\documentstyle[epsfig]{aipproc}

\begin{document}
\title{Contact Interactions and high $Q^2$ Events at HERA
\footnote{Talk given at the {\it 5th International Workshop on Deep 
Inelastic Scattering and QCD}, Chicago, Illinois, 14--18 April 1997.}
}
\author{Dieter Zeppenfeld}
\address{Physics Department, University of Wisconsin, Madison, WI 53706}

\maketitle

\begin{abstract}
Effective $eeqq$ contact interactions can enhance event rates 
in $e^+p$ scattering at HERA at high $Q^2$. Present constraints from atomic 
parity violation measurements and from Drell Yan events at the Tevatron are 
discussed.
\end{abstract}

Both the H1\cite{H1} and ZEUS\cite{zeus} experiments at HERA have observed an 
excess above SM expectations in $e^+p\to e^+jX$ events  at 
$Q^2>15{,}000\rm~GeV^2$. Only the H1 data, however, are suggestive 
of an $s$-channel (leptoquark) resonance~\cite{leptoq}, and other 
possible sources of rate 
enhancements in high $Q^2$ DIS events need to be investigated.  

A fairly model independent parameterization of possible new physics effects 
is provided by an effective Lagrangian of $eeqq$ contact 
interactions~\cite{altarelli,BCHZ}: any 
exchange, in the $s$-, $t$- or $u$-channel, can be cast into this form provided
the exchanged quanta are heavy compared to the center of mass energy of the 
scattering process. Following Ref.~\cite{BCHZ} the contact terms are 
parameterized as
\begin{eqnarray}
L_{NC} &=& \sum_q \Bigl[ \eta_{LL}\left(\overline{e_L} \gamma_\mu e_L\right)
\left(\overline{q_L} \gamma^\mu q_L \right) + \eta_{RR} \left(\overline{e_R}
\gamma_\mu e_R\right) \left( \overline{q_R}\gamma^\mu q_R\right) \nonumber\\
&& \qquad {}+ \eta_{LR} \left(\overline{e_L} \gamma_\mu e_L\right)
\left(\overline{q_R}
\gamma^\mu q_R\right) + \eta_{RL} \left(\overline{e_R} \gamma_\mu e_R\right)
\left(\overline{q_L} \gamma^\mu q_L \right) \Bigr] \,, \label{effL}
\end{eqnarray}
where the coefficients have dimension (TeV)$^{-2}$ and are conventionally
expressed as $\eta_{\alpha\beta}^{eq} = \pm 4\pi /\Lambda_{eq}^2$, with an 
effective mass scale of new physics, $\Lambda_{eq}$.

The contact terms (\ref{effL}) give additional contributions to the reduced 
amplitudes for $eq\to eq$, $\bar qq\to e^+e^-$, and $e^+e^-\to \bar qq$  
subprocesses, beyond the SM photon- and $Z$-exchange terms:
\begin{equation}
M_{\alpha\beta}^{eq} = {e^2 Q_e Q_q\over\hat t} + {g_Z^2 (T_{e\alpha}^3 - 
s^2_{\rm w}Q_e) (T_{q\beta}^3 - s^2_{\rm w} Q_q) \over \hat t - m_Z^2} +
\eta_{\alpha\beta}^{eq} \,. \label{amps}
\end{equation}
Here $Q_f$ and $T_{f\alpha}^3$ are charges and weak isospins, respectively, of
the external fermions $f_\alpha$, $g_Z = e/(\sin\theta_{\rm w} \, 
\cos\theta_{\rm w})$, and $s_{\rm w}=\sin\theta_{\rm w}$. 
The Mandelstam invariant $\hat t$ is given by
$\hat t = (E_{\rm c.m.})^2$ for $e^+e^-$, by $\hat t = -Q^2 = -sxy$ for $ep$
and by $\hat t  = sx_1x_2$ for $\ p\bar p$ collisions. 
Thus the presence of a contact interaction has interconnected implications for
$ep\to ejX$, $p\bar p\to e^+e^-X$, and $e^+e^-\to \rm hadrons$ and also for
atomic physics parity violation experiments.

\begin{figure} 
\centerline{\epsfig{file=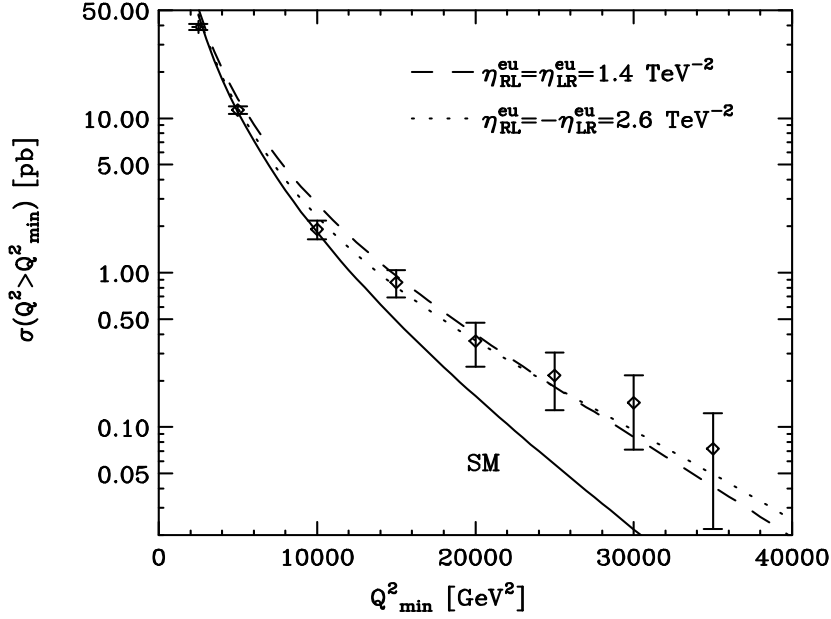,height=3.2in,width=4.3in}}
\vspace{10pt}
\caption{Integrated cross sections versus a minimum $Q^2$ for $e^+p\to e^+X$
for the SM (solid curve) and two choices of contact interactions.
The data points correspond to combined H1~\protect\cite{H1} and 
ZEUS~\protect\cite{zeus} measurements, assuming constant detection 
efficiencies of 80\% and 81.5\%, respectively.
Errors are statistical only. From Ref.~\protect\cite{BCHZ}.
}
\label{fig1}
\end{figure}

In $ep$ scattering at high $Q^2$ valence quark contributions dominate and the 
cross sections are related to the amplitudes (\ref{amps}) and parton 
distribution functions $q(x,Q^2)$ as
\begin{equation}
{d\sigma(e^+p)\over dx\,dy} \approx {sx\over16\pi} \sum_{q=u,d} q(x,Q^2)
\left[
\left| M_{LR}^{eq} \right|^2 + \left| M_{RL}^{eq} \right|^2 + (1-y)^2 \left(
\left| M_{LL}^{eq} \right|^2 + \left| M_{RR}^{eq} \right|^2 \right) \right]
\, .  \label{bar-ep}
\end{equation}
From (\ref{bar-ep}) we see that a high-$y$ anomaly in $e^+p$ requires 
an $\eta_{LR}^{eq}$ or $\eta_{RL}^{eq}$ amplitude. $\eta_{LL}^{eq}$ or 
$\eta_{RR}^{eq}$ amplitudes are suppressed by the $(1-y)^2$ factor 
in $e^+p$ collisions but they would be enhanced compared to
$\eta_{LR}$ or $\eta_{RL}$ terms in $e^-p$ scattering.
Because the $d$-parton density is severely suppressed in the large $x$ region
relative to the $u$-parton density,
the $eu$ contact interaction is needed to achieve an $e^+p$ cross section
enhancement. 

As an illustration consider two scenarios, 
$\eta_{RL}^{eu} =\eta_{LR}^{eu} =  1.4\rm\ TeV^{-2}$ and 
$\eta_{RL}^{eu} = -\eta_{LR}^{eu} =  2.6 \rm\ TeV^{-2}$, 
corresponding to scales $\Lambda_{eu}\approx 3$~TeV and
$\Lambda_{eu}\approx 2.2$~TeV, respectively. The resulting DIS cross 
sections, above a minimum $Q^2$, are shown in Fig.~1. Both scenarios 
provide for a better description of the HERA data than the SM. However, 
their consistency with other data must be investigated.

The choice $\eta_{RL}^{eu} = -\eta_{LR}^{eu}$ is parity violating and 
modifies the SM prediction for
the atomic physics parity violating parameter $Q_W$, given by
\begin{equation}
Q_W = Q_W^{\rm SM} - {1\over \sqrt{2}G_F} \left[ (N+2Z)\Delta\eta^{eu} +
(2N+Z)\Delta\eta^{ed} \right] \,.
\end{equation}
Here $N$ is the number of neutrons, $Z$ is the number of protons, and
$\Delta\eta^{eq}=\eta_{RR}^{eq}-\eta_{LL}^{eq}+\eta_{RL}^{eq}-\eta_{LR}^{eq}$
is the coefficient of the $VA$ combination of fermion currents which is probed.
The ${}^{133}_{\phantom055}$Cs measurements find $Q_W=-72.1\pm 0.9$~\cite{apv}
compared to a SM prediction of $Q_W^{\rm SM} = -72.9$~\cite{pdg} which yields 
\begin{equation}
\Delta\eta^{eu}+1.12\Delta\eta^{ed} = (-0.068\pm 0.082)\; {\rm TeV}^{-2}\;.
\end{equation}
The $\eta_{RL}^{eu} = -\eta_{LR}^{eu}$ scenario above is thus ruled out 
unless one introduces additional, cancelling terms such as 
$\eta_{LL}^{eu} = -\eta_{RR}^{eu}=\eta_{RL}^{eu}$. 
 
Such additional terms would only marginally affect the HERA 
observables, but they would give rise to even larger deviations from 
the SM in $p\bar p\to \ell^+\ell^-X$
at the Tevatron, in particular at large $\ell^+\ell^-$ invariant mass, where
contact interactions dominate the amplitudes and interference effects 
can be neglected. Recently, the CDF Collaboration has determined new 
preliminary limits on $\eta_{LL}^{eq}$ contact terms 
corresponding to $\Lambda_{eq}>3.4$~(2.4)~TeV for $\eta_{LL}^{eu}<0$ 
($\eta_{LL}^{eu}>0$)~\cite{bodek}. Comparison of the measured  
$\ell^+\ell^-$ invariant mass spectrum with predictions in the presence 
of contact terms indicate that the scenario 
$\eta_{RL}^{eu} =\eta_{LR}^{eu} =  1.4\rm\ TeV^{-2}$ above is still 
(marginally) compatible with the present Tevatron data while scenarios such as
$\eta_{RL}^{eu} =-\eta_{LR}^{eu} =  2.6\rm\ TeV^{-2}$ are clearly ruled out.
Present Tevatron Drell-Yan data probe contact terms at the same scale 
$\Lambda_{eq}$ as HERA DIS data, but the two machines are complementary in 
their sensitivity to particular combinations of couplings.

\section*{Acknowledgments}
I would like to thank V.~Barger, Kingman Cheung and K.~Hagiwara for the 
enjoyable collaboration which lead to the results presented in this talk.
This research was supported in part by the U.S.~Department of Energy under
Grant No. DE-FG02-95ER40896 and in part by the University of Wisconsin Research
Committee with funds granted by the Wisconsin Alumni Research Foundation.

\end{document}